\documentclass[11pt]{article}
\usepackage[a4paper,margin=2.5cm]{geometry}
\usepackage[T1]{fontenc}
\usepackage{lmodern}
\usepackage{amsmath,amssymb,amsthm}
\usepackage{graphicx}
\usepackage{booktabs}
\usepackage{makecell}
\usepackage[section]{placeins}
\usepackage{pifont}
\usepackage{worldflags}
\usepackage[numbers,sort&compress]{natbib}
\usepackage[hidelinks,hyperfootnotes=false]{hyperref}

\theoremstyle{definition}
\newtheorem{definition}{Definition}[section]
\newtheorem{example}[definition]{Example}

\title{Post-Pruning vs. Early Stopping: Which Works Better for Link Prediction with AnyBurl?}
\author{%
  Thomas Eiter\\
  TU Wien, Vienna, Austria\\
  \texttt{thomas.eiter@tuwien.ac.at}
  \and
  Davide Soldà\textsuperscript{*\,\ding{41}}\\
  TU Wien, Vienna, Austria\\
  \texttt{davide.solda@tuwien.ac.at}
  \and
  Hai Xia\textsuperscript{*\,\ding{41}}\\
  TU Wien, Vienna, Austria\\
  \texttt{hxia@ac.tuwien.ac.at}%
}
\date{}

\begin{document}
\maketitle
{\renewcommand{\thefootnote}{}%
\footnotetext{\textsuperscript{*}Both authors contributed equally to this research.\quad
\textsuperscript{\ding{41}}Corresponding authors.}}

\begin{abstract}
Rule-based link prediction combines symbolic and statistical artificial intelligence methods by inductively learning a set of rules from a knowledge graph (KG), with the aim of completing the graph when a query cannot be answered from the KG alone.
Several systems have been proposed, and while they perform well on classical benchmarks, even compared with embedding-based systems, many of them produce a large number of rules. This compromises scalability in the query-answering phase, especially under strong hardware constraints. In this work, we investigate the behavior of AnyBurl, a state-of-the-art rule learner for link prediction, to determine whether running it for a given time and then pruning the learned rules yields better query-answering performance at test time than a simpler approach that stops rule generation earlier.
\end{abstract}

\section{Introduction}
Knowledge Graphs (KGs) \cite{hogan2021knowledge} have attracted
increasing attention in recent years. For our purposes, a KG can be
regarded as a large collection of subject-relation-object triplets, for
instance $(\texttt{Bob}, \texttt{lives}, \texttt{London})$;
well-known examples include FreeBase \cite{bollacker2008freebase},
DBPedia \cite{auer2007dbpedia}, Microsoft Satori \cite{noy2019industry}, 
and YAGO \cite{suchanek2007yago}.
Although a KG typically gathers a large number of facts (triplets), it is
nonetheless assumed to be often incomplete \cite{dong2014knowledge}: 
entities, relations between entities, or both may be missing.
This observation raises the question of whether a graph can be
\emph{completed}. The problem has been capturing the 
interest of several communities, with applications in 
intelligent question answering
\cite{huang2019knowledge}, recommender systems
\cite{wang2019explainable} and network security systems
\cite{jia2018practical}.
Concretely, given a KG $\mathcal{G}$, a completion task amounts to
filling gaps such as \texttt{(Einstein, bornIn, ?)} or
\texttt{(?, worksAt, Apple)}. The most widespread approaches rely on
graph embeddings \cite{ali2021bringing,rossi2021knowledge,ruffinelli2020you} or, more recently, on large
language models \cite{shu2024knowledge}.
A parallel line of work instead learns symbolic representations
\cite{galarraga2013amie,galarraga2015fast,meilicke2018fine,
meilicke2019anytime,GaoICW24}: rules are induced from the data and link
prediction queries are then answered by applying these rules over the
knowledge graph, treated as a set of facts.
Closely related are neuro-symbolic methods, which come in various
forms. One such form injects logical knowledge (e.g.\ Description
Logics) into an embedding space by encoding constraints directly in the
objective function, as done in \cite{kulmanov2019embeddings,
abboud2020boxe,gutierrez2018knowledge,lourencco2026boxlite}.

In this work we focus on purely symbolic approaches. Specifically, we address the
problem of \emph{pruning} a set of learned rules, that is, of retaining only a
subset of them, say $n\%$ of the original set. Rule learners targeting link
prediction tend to produce very large rule sets, often in the order of millions of
rules. Such volume compromises (i) \emph{scalability and responsiveness}, since
every query must be answered by considering a vast number of rules;
(ii) \emph{inspectability}, since no human expert can realistically review the full
set of generated rules to decide which ones are worth retaining; and
(iii) \emph{usability as input to LLM-based pipelines}, since 
it is known in the literature that when LLM's context window are overloaded ---
with rules in our case, as shown in \cite{pesare26} --- they tend to perform worse.
Some other works identified the amount of rules
generated as a challenge for the problem of rule aggregation~\cite{betz2022supervised} or for the generation and evaluation as 
in~\cite{solda2026incremental}.

Concretely, we build on metrics discussed in 
\cite{galarraga2013amie}---\emph{support}, 
\emph{confidence}, and \emph{PCA confidence}---in addition to a variant of 
support that splits the rule budget across relations as the basis for a
heuristic selection among the rules produced by the state-of-the-art rule learner
AnyBurl \cite{meilicke2024anytime}. We then assess how such a pruning affects link
prediction performance, comparing the pruned rule set both against the original one
and against a rule set of comparable size obtained by shortening the time budget of
the anytime AnyBurl algorithm. Therefore, it can be read as a way of 
testing whether an early stopping criterion can be more or less beneficial 
with respect to a post-generation pruning phase applied to a larger set of generated
rules. Even if we test only AnyBurl, we can apply this study to other 
systems producing Horn clauses for link prediction tasks. 

\subsection{Related Works}

The closest related work is \cite{galarraga2013amie}, where the same metrics that we
employ for post-hoc pruning are instead applied during mining, in order to restrict
the search space. There, pruning is essential: the refinement operators are complete
with respect to the chosen language bias, so that an unrestricted top-down
exploration would, on the one hand, be computationally infeasible and, on the other,
return a large number of redundant rules.

More generally, pruning AI systems after training is a well-established practice.
In particular, a variety of approaches have been proposed for pruning neural
networks, both at the level of individual weights (unstructured pruning) and at the
level of entire neurons, channels or layers 
(structured pruning) \cite{cheng2024survey}. 
In this area, one of the most widely used heuristics is \emph{magnitude pruning},
which discards the weights with smallest absolute value. As in that setting, our
goal is to retain only a small fraction of the model while preserving predictive
performance. The difference lies in the units being pruned: ours are logical rules,
rather than weights, which are not individually interpretable. Therefore, 
post-training rule pruning could be considered as the symbolic counterpart of 
magnitude pruning and model compression.

\section{Background}

In this section, we recall some
definitions and 
notions regarding the link prediction problem. We also give
a brief introduction to the rule-based systems for link 
prediction that we consider for rule pruning.

\subsection{Link Prediction in  Knowledge Graphs}

Following \cite{liang2024survey}, we can formalize the notion of 
KG as:
\begin{definition}[KG]
A knowledge graph is defined as:
\[
\mathcal{G}
=\{(\texttt{a},\texttt{r},\texttt{b})\mid \texttt{a},\texttt{b}\in\mathcal{E},\; \texttt{r}\in\mathcal{R}\}
\subseteq \mathcal{E}\times\mathcal{R}\times\mathcal{E},
\]
where $\mathcal{E}$, $\mathcal{R}$ represent the sets of entities and
relations. Facts are in  a triplet format $(\texttt{a},\texttt{r},\texttt{b})$, 
where $\texttt{a}$, $\texttt{b}$, $\texttt{r}$
represent the head entity, the tail entity and the relation between them,
respectively.
\end{definition}
Since a KG $\mathcal{G}$ is inherently incomplete, the \emph{link prediction} 
(or \emph{knowledge graph completion})
task aims at inferring missing triplets from the observed ones. Formally,
\begin{definition}[Link Prediction]
Given an incomplete triplet, a \emph{query} takes one of the two forms
\[
q = (\texttt{a},\texttt{r},\texttt{?}) \qquad \text{or} \qquad 
q = (\texttt{?},\texttt{r},\texttt{b}),
\]
called \emph{tail query} and \emph{head query} respectively. 
Answering $q$ amounts to identifying the entity 
$\texttt{e}\in\mathcal{E}$ s.t. the completed 
triplet $(\texttt{a},\texttt{r},\texttt{e})$ (resp.\ 
$(\texttt{e},\texttt{r},\texttt{b})$) is a valid fact of the 
domain, although it may be absent from $\mathcal{G}$.
\end{definition}

In practice, the task is cast as a \emph{ranking} problem. A model learns a
scoring partial function
\[
f_{\theta}:\ \mathcal{E}\times\mathcal{R}\times\mathcal{E}\longrightarrow\mathbb[0,1],
\]
parameterised by $\theta$, 
which assigns to each candidate triplet a
plausibility score, with higher values indicating higher likelihood of the
triplet being true. $f_{\theta}$ is a partial function, as it 
might not associate a score to certain facts.
Given a query $(\texttt{a},\texttt{r},\texttt{?})$, every entity
$\texttt{e}\in\mathcal{E}$ is scored through $f_{\theta}(\texttt{a},
\texttt{r},\texttt{e})$ and the candidates
are sorted in decreasing order of their score; the head query 
$(\texttt{?},\texttt{r},\texttt{b})$ is
handled symmetrically via $f_{\theta}(\texttt{e},
\texttt{r},\texttt{b})$. The {\em rank}\/ of the ground-truth
entity is then
\[
\begin{aligned}
&\mathrm{rank}(\texttt{b} \mid \texttt{a},\texttt{r})
= 1 + \bigl|\{\, \texttt{e}\in\mathcal{E}\setminus\{\texttt{b}\} \ :\
f_{\theta}(\texttt{a},\texttt{r},\texttt{e}) > f_{\theta}(\texttt{a},
\texttt{r},\texttt{b}) \,\}\bigr|,\\
&\mathrm{rank}(\texttt{a} \mid \texttt{r},\texttt{b})
= 1 + \bigl|\{\, \texttt{e}\in\mathcal{E}\setminus\{\texttt{a}\} \ :\
f_{\theta}(\texttt{e},\texttt{r},\texttt{b}) > f_{\theta}(\texttt{a},
\texttt{r},\texttt{b}) \,\}\bigr|;
\end{aligned}
\]
furthermore, in the standard \emph{filtered} setting, all entities $\texttt{e}\neq b$ such that
$(\texttt{a},\texttt{r},\texttt{e})$ already appears in the training, 
validation, or test set are
removed from the candidate set before ranking, as done in the
seminal work \cite{bordes2013translating}, 
so that other correct answers are not penalized. 
Model quality is assessed by rank-based metrics
such as {\em Mean Reciprocal Rank (MRR)}\/ and {\em Hits@$k$}, averaged over all head and
tail queries obtained from the evaluation triplets.

\begin{definition}[MRR and Hits@$k$]
Let $\mathcal{T}_{\mathrm{test}}$ be the set of evaluation triplets. Each
$(\texttt{a},\texttt{r},\texttt{b})\in\mathcal{T}_{\mathrm{test}}$ 
yields two queries, $(\texttt{a},\texttt{r},\texttt{?})$ and
$(\texttt{?},\texttt{r},\texttt{b})$, 
with filtered ranks $\mathrm{rank}(\texttt{b}\mid \texttt{a},\texttt{r})$ and
$\mathrm{rank}(\texttt{a}\mid \texttt{r},\texttt{b})$. Then
\begin{align*}
\mathrm{MRR}
= \ & \ \frac{1}{2|\mathcal{T}_{\mathrm{test}}|}
\sum_{(\texttt{a},\texttt{r},\texttt{b})\in\mathcal{T}_{\mathrm{test}}}
\left(
\frac{1}{\mathrm{rank}(\texttt{b}\mid \texttt{a},\texttt{r})}
+
\frac{1}{\mathrm{rank}(\texttt{a}\mid \texttt{r},\texttt{b})}
\right), \\
\mathrm{Hits}@k
= \ &\ \frac{1}{2|\mathcal{T}_{\mathrm{test}}|}
\sum_{(\texttt{a},\texttt{r},\texttt{b})\in\mathcal{T}_{\mathrm{test}}}
\Bigl(
\begin{array}{l}
\mathbf{1}\bigl[\mathrm{rank}(\texttt{b}\mid \texttt{a},\texttt{r})\le k\bigr]
+\\
\mathbf{1}\bigl[\mathrm{rank}(\texttt{a}\mid \texttt{r},\texttt{b})\le k\bigr]
\end{array}
\Bigr),    
\end{align*}
where $\mathbf{1}[\cdot]$ is the indicator function and $k\in\{1,3,10\}$
typically. Both metrics lie in $[0,1]$ and higher is better.
\end{definition}

Hits@$k$ is suited to use cases where it is sufficient for the correct answer
to appear among the top $k$ results. Unless $k=1$, there is no benefit in
ranking the correct entity first, as long as it appears among the top $k$
candidates.

In contrast, MRR is sensitive to the exact position
of the correct entity: if it is ranked at position $i+1$ instead of $i$, the
prediction is penalized. The respective penalty $\frac{1}{i} - \frac{1}{i+1} =
\frac{1}{i(i+1)}$ is largest at the top of the ranking and quickly becomes
negligible for lower positions, so MRR rewards ranking the correct answer
near the top rather than within a fixed cutoff.
As a consequence, when a short list of $k$ answers can be taken into consideration,
then evaluating a system against Hits@$k$ is a good choice, whereas in a scenario
where each answer inspection comes with a cost, then MRR is a better fit.

\subsection{ILP Systems for KG Link Prediction}
We first formally introduce the concepts of \emph{entailment}, \emph{rules}, 
\emph{derivation}. Then, we describe what is the ILP task for KG Link prediction,
and finally we introduce some standard metrics.
\paragraph{Entailment.}
Let $\mathcal{G}$ be a knowledge graph over entities $\mathcal{E}$ and relations
$\mathcal{R}$. For a ground atom we set $\mathcal{G} \models p(a,b)$ if and only if
$p(a,b) \in \mathcal{G}$. More generally, let $F$ be a conjunction of atoms with
variables $X_1, \dots, X_k$; then
\[
\begin{aligned}
&\mathcal{G} \models \exists X_1, \dots, X_k \,.\, F
  \quad\text{iff}\\
&\quad \text{there exists } \sigma : \{X_1, \dots, X_k\} \to \mathcal{E}
  \text{ such that } \mathcal{G} \models F\sigma,
\end{aligned}
\]
where $F\sigma$ denotes the set of ground atoms obtained by applying the
substitution $\sigma$ to $F$. Whenever $\mathcal{G}$ does not satisfy a formula $F$,
we write $\mathcal{G} \not\models F$.

\paragraph{Rules.}
Given a rule $r : h(H_1, H_2) \leftarrow b_1(B_1, B_2), \dots, b_m(B_m, B_{m+1})$
with $H_1 = B_i$ and $H_2 = B_j$ for some $i,j\in [1, m+1]$, 
as we require every variable occurring in the
head to occur in the body as well. Such a constraint correspond to the
\emph{safety} requirement of logic programming. Note that if $H_1=H_2$, then
$i$ may coincide with $j$.

\paragraph{Derivation.}
We say that $\mathcal{G}$ \emph{derives} $a$ for the query $q(X, b)$ with score
$s_r$ if there exists a rule $r$ with score $s_r$, head $q(H_1, H_2)$ and body
$b_1(B_1, B_2), \dots, b_m(B_m, B_{m+1})$ such that
\[
\begin{aligned}
&\mathcal{G} \models \exists B_1, B_2, \dots, B_m \,.\,
  b_1(B_1, B_2) \wedge \dots \wedge b_m(B_m, B_{m+1}),\\
&\quad \text{with } B_i = a,\ B_j = b  
\end{aligned}
\]
The derivation as a second argument of the $q$-relation is defined symmetrically.

Therefore, the ILP task for the link prediction problem can be
formulated as the inductive generation of rules $R$ to be able to
perform the link prediction task, meaning that once we have a query
$\texttt{q}(\texttt{?}, \texttt{b})$ or $\texttt{q}(\texttt{a}, \texttt{?})$,
then we list out as answers to the query as a list of constants labeled with
their score. If there are different rules that we can derive a specific constant
$e$, then we label such a constant with the maximum among those scores. 
We thus define the score associated with $e$ for the query $q(X, e')$ as:
\[
\begin{aligned}
f_\theta(e, q, e')\ :=\ \max\{&s_r \text{ rule }r \text{ deriving } e
\text{ with second }\\ 
&\qquad \text{ argument instantiated with } e'\}    
\end{aligned}
\]
where $s_r$ is the score associated to the rule $r$. The symmetric case
of $e$ being the second argument in the query is symmetrically defined.

The support is formally defined as follows:
\[
\mathrm{supp}(r)=
\left|\left\{ (e, h, e')  \ :\
       \begin{array}{l}
         \mathcal{G} \models
           \exists B_1, \dots, B_{m+1}\ . \\[2pt]
         h(e, e') \bigwedge_{i=1, \dots, m} b_i(B_i, B_{i+1})
       \end{array}
     \right\}\right|
\]
where $r$ is the rule 
$h(H_1, H_2) \leftarrow b_1(B_1, B_2), \dots, b_m(B_m, B_{m+1})$ and
when $H_1=B_i$ and $H_2=B_j$, then $B_i$ is replaced with $e$ and 
 $B_j$ with $e'$, respectively.

The (standard) confidence is formally defined as follows:
\[
\mathrm{conf}(r)=
\frac{\mathrm{supp}(r)}
     {\left|\left\{  
     (e, h, e') \ :\
       \mathcal{G} \models \exists B_i \bigwedge_{i=1, \dots, m} b_i(B_i, B_{i+1})
     \right\}\right|}
\]
where $r$ is the rule 
$h(H_1, H_2) \leftarrow b_1(B_1, B_2), \dots, b_m(B_m, B_{m+1})$ and where $B_i=e$ if $B_i=H_1$
and $B_j=e'$ if $B_j=H_2$.

The confidence of rules is therefore the result of dividing the number of 
groundings for the conjunction of
body and head by the number of body groundings. In this way we can normalize 
the number of representatives (supports) the rule stands for with the number of
times the rule is actually triggered.
\paragraph{AnyBurl}

AnyBurl, as presented in \cite{meilicke2024anytime}, is a random-walk\\ 
reinforcement-based rule learner. From a semantical point of view, it 
uses the \emph{object identity assumption} \cite{semeraro1994avoiding}, 
meaning that different variables $X$ and $Y$ can be grounded only with different 
entities/constants $e$ and $e'$, where $e\neq e'$, i.e. the variables appearing
in the body of a rule are meant to satisfy an all different constraint. 
Rules are generated by executing random walks on
a graph. Such paths are generalized into rules by replacing constants with variables. 
Most rules produced by AnyBurl belong to one of the following three cases:
\begin{itemize}
    \item binary rules \textbf{B}: 
    \[
    h(A_0, A_n) \leftarrow \bigwedge\limits_{i=1}^n b_i(A_{i-1}, A_i).
    \]
    \item unary rules with dangling atom \textbf{U}$_\textbf{d}$:
    \[
    h(A_0, c) \leftarrow \bigwedge\limits_{i=1}^n b_i(A_{i-1}, A_i).
    \]
    \item unary rules with an atom containing a constant \textbf{U}$_\textbf{c}$:
    \[
    h(A_0, c) \leftarrow \bigwedge\limits_{i=1}^{n-1} b_i(A_{i-1}, A_i)\ \land\ b_n(A_{n-1},c').
    \]
\end{itemize}
In addition to them, each of the patters may contain atoms where the arguments are 
swapped, for instance if $h(A_0, c) \leftarrow b(A_0, A_1).$ is a \textbf{U}$_\textbf{d}$
rule, then also $h(c, A_0) \leftarrow b(A_0, A_1).$, $h(c, A_0) \leftarrow b(A_1, A_0).$
and $h(A_0, c) \leftarrow b(A_1, A_0).$ are \textbf{U}$_\textbf{d}$ rules.
AnyBURL also learns \textbf{Z} rules with an empty body, 
$h(A_0, c) \leftarrow \top$ or $h(c, A_0) \leftarrow \top$, which predict the 
constant $c$ whatever the query entity, and rules whose head relates an entity to 
itself, $h(A_0, A_0) \leftarrow b(A_0, A_1)$. Note that only the 
\textbf{Z} rules do not satisfy the condition
of being closed rules. We later prune the \textbf{U}$_\textbf{c}$ rules out in case they
do not generalize on the training set.

Interestingly for our study, \cite{meilicke2024anytime} proposes an original 
confidence mechanism for scoring rules based on a sampling mechanism; 
such an approach consists in sampling walks on a graph such that they are
consistent with a given first-order rule's body. If the walk-induced grounding yields
to a ground true head atom, then the rule is considered to predict
such an atom correctly. Let us consider the following example.
\begin{example}   
Given a rule 
\[speaks(A_0, A_2) \leftarrow livesIn(A_0, A_1), isSpokenIn(A_2, A_1).,\]
the 
search starts by collecting all the possible grounding of the variable $A_0$
such that they appear as the first argument of a $speaks$-relation, for 
instance $\{\texttt{Alice}, \texttt{Bob}, \texttt{Charlie}\}$. A random element such as
\texttt{Bob} is picked, then a random edge labeled with $speaks$ 
connecting \texttt{Bob} to another entity $e$ is randomly
chosen. If such an entity $e$ is $\texttt{Dutch}$, then an entity like
\texttt{Netherlands} from which there is an outgoing arc labeled with $isSpokenIn$ 
ending up into $\texttt{Dutch}$ may be chosen. If the triplet $(\texttt{Bob}, speaks, 
\texttt{Dutch})$ belongs to the KG, then it counts as a contribution to both the numerator
and the denominator of the confidence score, otherwise only to the denominator. 
\end{example}

Since the generation of rules happens by generalizing paths induced by
random walks on the graph, AnyBurl classifies the random walks into profiles, where a
profile consists of the type of rule (\textbf{B}, \textbf{U}$_\textbf{d}$ or \textbf{U}$_\textbf{c}$) 
and of the length of the path considered. In order to enhance the decision on 
how much CPU should be allocated to a specific profile, a RL policy is used, whose
rewards consist of some confidence based metrics. 

\paragraph{AMiE}

AMiE was first introduced in \cite{galarraga2013amie}. One of the language biases that is 
imposed here with respect to AnyBurl is the fact that only closed rules are learned, 
which rules out the $\mathbf{U}_\mathbf{d}$ rule-schema. Another difference with respect
to AnyBurl is that the rules learned by AMiE do not fulfill the path constraint as 
$\mathbf{B}$, $\mathbf{U}_\mathbf{c}$, and $\mathbf{U}_\mathbf{d}$ rule-schema do.
The algorithm proceeds by applying a set of operations called $\mathcal{O}_\text{D}$, 
$\mathcal{O}_\text{I}$, and $\mathcal{O}_\text{C}$. More precisely,
\begin{itemize}
    \item $\mathcal{O}_\text{D}$ takes a rule as an input and returns a rule with
    a body extended with a new atom with two variables as argument which shares one variable
    with the ones already present in the rule;
    \item $\mathcal{O}_\text{I}$ takes a rule as an input and returns a rule with
    a body extended with a new atom with one variable that already appears in the rule 
    as one argument, while the other argument is an entity from $\mathcal{E}$;
    \item $\mathcal{O}_\text{C}$ takes a rule as an input and returns a rule with
    a body extended with a new atom with two variables already appearing in the rule.
\end{itemize}
Non-closed rules are never output, but a later refinement may close them.
We first define the concept of \emph{head coverage}, and then introduce AMiE's algorithm.
The head coverage of a rule $r$ is obtained as the ratio between the support of $r$ and
the number pairs connected via the head predicate name in the KG. Formally, let
$r$ be 
$$h(H_1, H_2) \leftarrow b_1(B_1, B_2), \dots, b_m(B_m, B_{m+1})$$ 
then
\[
\begin{aligned}
&\mathrm{head\_coverage}(r) = \frac{\mathrm{supp}(r)}{|\{(e, h, e')\ :\ \mathcal{G}\models h(e, e')\}|}
\end{aligned}
\]
The idea is to normalize the support of a rule by dividing it by the 
number of entity pairs connected by the head predicate in the KG.

The rule-generation AMiE algorithm proceeds as follows: 
\begin{enumerate}
    \item initialize a FIFO queue $q$ with an empty list as the only element for a certain
    head $r(X, Y)$;
    \item while $q$ contains at least one element;
    \begin{enumerate}
        \item extract a rule $r$ from $q$;
        \item if $r_p=\mathrm{parent}(r)$ is defined, $r$ is closed,\\ and $\mathrm{conf}(r_p)<\mathrm{conf}(r)$, then output $r$;
        \item for all operators $o$, for all rules $r'$ obtained from $o(r)$:
        \begin{enumerate}
            \item if $r'$ is not pruned then add it to the queue $q$;
        \end{enumerate}
    \end{enumerate}
\end{enumerate}
where the operator $o$ is the one that can be obtained by instantiating the various
$\mathcal{O}_\text{D}$, $\mathcal{O}_\text{I}$, and $\mathcal{O}_\text{C}$. The pruning
check is passed if the following two conditions are matched: (i) refinement's  
head coverage is at least a threshold ($0.01$ in~\cite{galarraga2013amie}) 
(ii) no rule $r''=r'$
was already produced with the same head coverage and confidence score. 

\begin{table}[t]
\centering
\caption{\rm Statistics of the benchmark datasets.}
\label{tab:datasets}
\small
\begin{tabular}{lrrrrr}
\toprule
Dataset & Entities & Relations & Train & Validation & Test  \\
\midrule
WN18RR     &    40,943 &  11 &     86,835 &  3,034 &  3,134  \\
FB15k-237  &    14,541 & 237 &    272,115 & 17,535 & 20,466  \\
YAGO3-10   &   123,182 &  37 &  1,079,040 &  5,000 &  5,000  \\
CoDEx-S    &    2,034  &  42 &    32,888  & 1,827  & 1,828   \\
CoDEx-M    &    17,050 &  51 &    185,584 & 10,310 & 10,311  \\
CoDEx-L    &    77,951 &  69 &    551,193 & 30,622 & 30,622  \\
Hetionet   &    45,158 &  24 &  2,249,925 &    121 & 151     \\
Nations    &        14 &  55 &      1,592 &    199 & 201     \\
\bottomrule
\end{tabular}
\end{table}

AMiE scores the rules it produces according to the PCA confidence. This measure was
introduced in \cite{galarraga2013amie} and aims at rectifying the standard confidence
measure, which is inadequate under the Open World Assumption (OWA), which is the working
assumption in this setting, as it counts every predicted fact that is absent from
the KB as a counterexample. The PCA confidence instead relies on the Partial
Completeness Assumption: if the KB contains at least one object for a given subject and
relation, it is assumed to contain all of them. For instance, if the KB contains the
triple $(\texttt{Alice}, \texttt{owns}, \texttt{house}_1)$, then it is assumed to
contain every true triple of the form $(\texttt{Alice}, \texttt{owns}, x)$; consequently,
any such triple that is \emph{not} in the KB is taken to be false and may serve as a
counterexample. Conversely, if the KB contains no triple
$(\texttt{Bob}, \texttt{owns}, x)$ at all, then nothing is assumed about what
\texttt{Bob} owns, and no counterexample is derived for him.
The PCA confidence is formally defined as follows:
\[
\mathrm{pca\_conf}(r)=
\frac{\mathrm{supp}(r)}
     {\left|\left\{ (e, h, e') \in \mathcal{G}\,{:}\ \ \
       \begin{array}{@{}l@{~}l@{}}
         \mathcal{G} \models &
           \exists B_1, \dots, B_{m+1}, H_2 \ . \\[2pt]
         & h(e, H_2) \bigwedge_{i=1}^m b_i(B_i, B_{i+1})
       \end{array}
     \right\}\right|}
\]
where $r$ is the rule 
$h(H_1, H_2) \leftarrow b_1(B_1, B_2), \dots, b_m(B_m, B_{m+1})$.

\section{Pruning Method}

Given a set of Horn clauses $R$ produced by AnyBurl, we present in what follows
some selection criterion $s$ such that given the training KG 
$\mathcal{G}_\text{train}$ and a value $p\in [0, 1]$, it returns a subset
$R'$ of $R$ such that $|R'|=\lfloor p \cdot |R| \rfloor$.

\begin{itemize}
    \item \textbf{Support}: the number of training triples a rule predicts correctly, as estimated by AnyBURL;
    \item \textbf{Confidence}: which is %
    the ration between the support and
    the body support;
    \item \textbf{Balanced}: recounted exact support values, with the rule budget split across head relations by D’Hondt apportionment (see below).;
    \item \textbf{PCA}: the PCA confidence of AMIE (Section 2.2), counted on
    the training graph.
\end{itemize}

We introduced the \textbf{balanced} criterion because a query with a given
relation name can only be answered by rules whose head atom features that same
relation. Rules with different head relations therefore serve disjoint sets of
queries, and they can be pruned independently. Building on this observation, we
allocate the rule budget using D’Hondt apportionment \cite{balinski2010fair}, 
where the ``votes'' are meant to be the number of occurrences of validation 
triples with that relation, and each relation fills its ``seats'' with its rules 
of highest support computed on the training set.\footnote{For a few
hundred rules on CoDEx-M, CoDEx-L, and YAGO3-10 whose predictions are too many to enumerate,
the count is estimated from a sample.}

\section{Experimental Evaluation}

We tested our methodology on the following standard benchmarks, whose
statistics are reported in Table~\ref{tab:datasets}:
\begin{itemize}
    \item WN18RR~\cite{dettmers2018convolutional}, which contains lexical and
    semantic information about the English language;
    \item FB15k-237~\cite{toutanova2015observed}, which contains cross-domain
    real-world facts about entities such as people, locations, movies, sports
    teams, and organizations;
    \item YAGO3-10~\cite{mahdisoltani2014yago3}, which contains information
    about people, places, and organisations;
    \item CoDEx-\{S,M,L\}~\cite{safavi2020codex}, which contains general knowledge about
    world history, culture, and science extracted from Wikidata and Wikipedia;
    \item Hetionet~\cite{himmelstein2017systematic}, which is an integrated 
    knowledge graph built by unifying data extracted from 29 public biomedical 
    databases;
    \item Nations~\cite{kemp2006learning}, which contains international relations 
    and geopolitical data.
\end{itemize}

Hetionet and Nations are an order of magnitude smaller than the rest in the number of test 
queries, so a single
query moves Hits@1 by 0.33 and 0.25 percentage points there; we still report 
the results concerning these two datasets, but we treat them as exploratory and
exclude them from statements made across benchmarks.

\begin{table}[t]%
\centering
\caption{\rm Analysis of the PCA metrics}
\label{tab:coverage}
\footnotesize
\setlength{\tabcolsep}{5pt}
\begin{tabular}{lrrrr}
\toprule
 & \multicolumn{2}{c}{support} & \multicolumn{2}{c}{PCA confidence}\\
\cmidrule(lr){2-3}\cmidrule(lr){4-5}
 & cov. & RR\,$|$\,cov. & cov. & RR\,$|$\,cov.\\
\midrule
WN18RR               & 71.9\%  & 0.672 & 6.2\%  & \textbf{0.931}\\
FB15k-237            & 89.0\%  & 0.331 & 15.1\% & \textbf{0.606}\\
CoDEx-S              & 97.9\%  & 0.415 & 38.3\% & \textbf{0.753}\\
CoDEx-M              & 96.9\%  & 0.320 & 33.5\% & \textbf{0.580}\\
CoDEx-L              & 85.1\%  & 0.331 & 7.6\%  & \textbf{0.650}\\
YAGO3-10             & 94.0\%  & 0.532 & 53.9\% & \textbf{0.756}\\
Hetionet$^{\dagger}$ & 100.0\% & 0.332 & 24.5\% & \textbf{0.644}\\
Nations$^{\dagger}$  & 72.6\%  & \textbf{0.812} & 17.9\% & 0.697\\
\bottomrule
\end{tabular}
\end{table}

\begin{figure*}[h]
\centering
\caption{\rm AnyBurl run with different time budgets (snapshots).
}
\includegraphics[width=\textwidth]{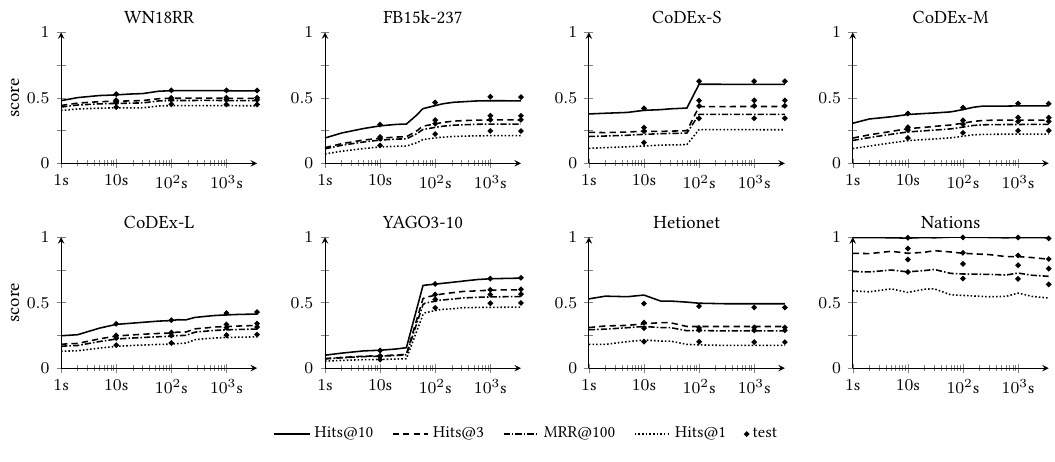}

\label{fig:metrics}
\end{figure*}

Given the set of rules produced by AnyBurl, 
we apply the four different filtering criteria given in Section 3. 
Table~\ref{tab:retention_hits} shows that the \textbf{support} criterion and its weighted
version 
\textbf{balanced} are effective metrics for the pruning of the rules.
As expected, by retaining 10\% of rules, we obtain better performance than
by retaining only 1\% of rules. Furthermore, we tested the performance on 
smaller set of rules (0.1\%, 0.25\%, 0.5\%), yielding the results in Table~\ref{tab:results-01}.
They show that if we admit even fewer rules, the pruning methods help in 
producing even more quality rules compared to the ones produced by AnyBurl if
a very short time budget is provided.
More precisely, in Figure~\ref{fig:metrics} we show the performance
of AnyBurl with different time budgets. Interestingly, it is often the case that
we need to produce a good amount of rules before discovering high quality rules.
Consequently, if the target set of rules is (too) small, then the pruning techniques
wins under \textbf{support} or \textbf{balanced} selection criteria 
over the simple earlier stopping approach.
Each panel follows one benchmark's rule set as AnyBurl is given more
time, from one second to one hour on a logarithmic scale, and reports the four quality measures of that rule
set. Note how most benchmarks rise steeply and then flatten well before the one hour time limit; %
on Hetionet and Nations, quality instead
drops after an early peak. Hetionet$^{\dagger}$ and Nations$^{\dagger}$ are exploratory, as these two 
benchmarks have only 151 and 201 test triples, respectively, making their results less reliable.

\begin{table}[h]
\centering
\caption{\rm Pruning against early stopping at three small target rule ratios:
0.1\%, 0.25\%, 0.5\%. %
\emph{Early stop} 
is the snapshot AnyBurl had written when it reached that size, \emph{Pruned} is the one-hour rule set cut to the same size, and both columns are MRR@100 on the test queries. 
\emph{Actual} is the fraction of rules effectively retained and \emph{Rules} their number. $\Delta = \mathrm{MRR}_{\text{pruned}} - \mathrm{MRR}_{\text{early stop}}$ for the support and the balanced criterion, so positive values favor pruning.  
Note that under the \emph{Early stop} and \emph{Pruned} columns, the MRR@100
obtained via the support metrics is displayed.
}
\label{tab:results-01}
\small
\setlength{\tabcolsep}{4pt}
\begin{tabular}{lrrrrrr}
\toprule
Dataset & Actual & Rules & \makecell[r]{Early\\stop} & Pruned & $\Delta$ support & $\Delta$ balanced \\
\midrule
\multicolumn{7}{l}{\emph{Target 0.1\,\%}} \\
YAGO3-10  & 0.103\% &  2,742 & 0.0552 & 0.4613 & $+0.4061$ & $+0.4289$ \\
WN18RR    & 0.091\% &     64 & 0.1310 & 0.4175 & $+0.2865$ & $+0.2914$ \\
CoDEx-S   & 0.110\% &  1,292 & 0.2474 & 0.3674 & $+0.1200$ & $+0.1332$ \\
CoDEx-L   & 0.094\% &  1,151 & 0.1291 & 0.2145 & $+0.0854$ & $+0.1100$ \\
CoDEx-M   & 0.102\% &  4,954 & 0.2354 & 0.2646 & $+0.0292$ & $+0.0357$ \\
FB15k-237 & 0.089\% &  5,338 & 0.1774 & 0.1999 & $+0.0225$ & $+0.0806$ \\
Nations$^{\dagger}$   & 0.110\% & 33,882 & 0.8415 & 0.4962 & $-0.3454$ & $-0.1834$ \\
\midrule
\multicolumn{7}{l}{\emph{Target 0.25\,\%}} \\
YAGO3-10     & 0.233\% &  6,194 & 0.0783 & 0.4948 & $+0.4165$ & $+0.4116$ \\
FB15k-237 & 0.284\% & 17,527 & 0.1461 & 0.2449 & $+0.0988$ & $+0.1270$ \\
CoDEx-S   & 0.257\% &  3,023 & 0.3462 & 0.3843 & $+0.0381$ & $+0.0519$ \\
CoDEx-M   & 0.263\% & 14,732 & 0.2709 & 0.2849 & $+0.0139$ & $+0.0187$ \\
\midrule
\multicolumn{7}{l}{\emph{Target 0.5\,\%}} \\
FB15k-237 & 0.552\% & 34,102 & 0.1739 & 0.2732 & $+0.0994$ & $+0.1144$ \\
CoDEx-L   & 0.496\% &  6,077 & 0.2106 & 0.2620 & $+0.0513$ & $+0.0508$ \\
CoDEx-M   & 0.427\% & 20,687 & 0.2719 & 0.2903 & $+0.0184$ & $+0.0221$ \\
YAGO3-10  & 0.544\% & 16,315 & 0.4920 & 0.4969 & $+0.0048$ & $-0.0007$ \\
\bottomrule
\end{tabular}
\end{table}

Interestingly, we can notice from Table~\ref{tab:retention_hits} that the 
\textbf{PCA confidence} does not score high on MRR and Hit@$k$ for 
$k\in \{1, 3, 10\}$. The reason we found is that this
metric scores high on rules when they can be fruitfully applied, but
the \emph{relevance} of a rule is not taken into consideration: 
a rule may have a very high PCA confidence, but it may be triggered only a handful of times. Such a conclusion is
also backed by Table~\ref{tab:coverage}, where 1\% of the rules is 
retained from pruning. In this table, cov.\ stands for the 
ratio of true answers that the pruned
rules are able to predict regardless of their ranks with respect to the total
number of test queries; RR\,$|$\,cov.\ is MRR@100 restricted
to the test queries for which there is a rule deriving them.

Furthermore, in order to simplify the MRR computation, we replaced MMR with
MRR@100, which computes the average of the reciprocal ranks over the pool of the
top 100 ranked elements, which is a common practice in the literature 
\cite{meilicke2020reinforced}. 
This yields an underestimate of the
actual MRR, but the difference is negligible. In fact, as observed above, 
the MRR penalty is larger at the top of the ranking and insignificant after a 
certain threshold.

\begin{table*}[h]
\caption{\rm Keeping 1\% and 10\% of rules according to support, 
balanced, standard confidence, and PCA criteria. $^{\dagger}$Hetionet has 151 test triples (302 queries), so its values are
exploratory.}
\label{tab:retention_hits}
\footnotesize
\setlength{\tabcolsep}{3.5pt}
\begin{tabular}{ll r rrrr rrrr}
\toprule
 & & & \multicolumn{4}{c}{keeping $1\,\%$ of the rules} & \multicolumn{4}{c}{keeping $10\,\%$ of the rules}\\
\cmidrule(lr){4-7}\cmidrule(lr){8-11}
 & & full & support & balanced & std.\ conf. & PCA & support & balanced & std.\ conf. & PCA\\
\midrule
WN18RR               & MRR@100 & 0.4957 & \textbf{0.4834} & 0.4802 & 0.3555 & 0.0576 & \textbf{0.4905} & 0.4879 & 0.3962 & 0.4074\\
                     & Hits@1  & 0.4568 & \textbf{0.4469} & 0.4446 & 0.3527 & 0.0563 & \textbf{0.4517} & 0.4488 & 0.3899 & 0.4001\\
                     & Hits@3  & 0.5105 & \textbf{0.4981} & 0.4935 & 0.3567 & 0.0586 & \textbf{0.5043} & 0.5027 & 0.4003 & 0.4119\\
                     & Hits@10 & 0.5715 & \textbf{0.5571} & 0.5509 & 0.3618 & 0.0600 & \textbf{0.5678} & 0.5641 & 0.4094 & 0.4223\\
\addlinespace[2pt]
FB15k-237            & MRR@100 & 0.3323 & 0.2950 & \textbf{0.3036} & 0.1060 & 0.0919 & 0.3235 & \textbf{0.3267} & 0.2456 & 0.2608\\
                     & Hits@1  & 0.2466 & 0.2207 & \textbf{0.2263} & 0.0950 & 0.0866 & 0.2410 & \textbf{0.2438} & 0.2005 & 0.2159\\
                     & Hits@3  & 0.3633 & 0.3214 & \textbf{0.3306} & 0.1136 & 0.0934 & 0.3528 & \textbf{0.3558} & 0.2676 & 0.2842\\
                     & Hits@10 & 0.5065 & 0.4445 & \textbf{0.4576} & 0.1263 & 0.1018 & 0.4913 & \textbf{0.4949} & 0.3350 & 0.3464\\
\addlinespace[2pt]
CoDEx-M              & MRR@100 & 0.3197 & 0.3106 & \textbf{0.3144} & 0.1541 & 0.1943 & 0.3175 & \textbf{0.3191} & 0.2205 & 0.2505\\
                     & Hits@1  & 0.2488 & 0.2426 & \textbf{0.2447} & 0.1371 & 0.1742 & 0.2475 & \textbf{0.2489} & 0.1875 & 0.2146\\
                     & Hits@3  & 0.3490 & 0.3405 & \textbf{0.3444} & 0.1629 & 0.2038 & 0.3473 & \textbf{0.3485} & 0.2366 & 0.2685\\
                     & Hits@10 & 0.4563 & 0.4380 & \textbf{0.4448} & 0.1851 & 0.2311 & 0.4496 & \textbf{0.4534} & 0.2813 & 0.3161\\
\addlinespace[2pt]
YAGO3-10             & MRR@100 & 0.5658 & 0.5003 & \textbf{0.5030} & 0.4566 & 0.4072 & 0.5266 & 0.5239 & 0.5334 & \textbf{0.5389}\\
                     & Hits@1  & 0.4981 & 0.4274 & \textbf{0.4287} & 0.4219 & 0.3687 & 0.4574 & 0.4527 & 0.4804 & \textbf{0.4852}\\
                     & Hits@3  & 0.6036 & 0.5359 & \textbf{0.5419} & 0.4768 & 0.4301 & 0.5599 & 0.5580 & 0.5679 & \textbf{0.5737}\\
                     & Hits@10 & 0.6905 & 0.6405 & \textbf{0.6459} & 0.5242 & 0.4827 & 0.6590 & \textbf{0.6613} & 0.6296 & 0.6366\\
\addlinespace[2pt]
Hetionet$^{\dagger}$ & MRR@100 & 0.2830 & 0.3317 & \textbf{0.3517} & 0.0731 & 0.1577 & 0.2830 & \textbf{0.2834} & 0.2178 & 0.2661\\
                     & Hits@1  & 0.1921 & 0.2384 & \textbf{0.2748} & 0.0596 & 0.1424 & 0.1954 & \textbf{0.1987} & 0.1689 & 0.1954\\
                     & Hits@3  & 0.3146 & \textbf{0.3642} & \textbf{0.3642} & 0.0828 & 0.1656 & \textbf{0.3046} & 0.3013 & 0.2318 & \textbf{0.3046}\\
                     & Hits@10 & 0.4536 & 0.5166 & \textbf{0.5265} & 0.0993 & 0.1821 & 0.4404 & \textbf{0.4437} & 0.3113 & 0.3974\\
\bottomrule
\end{tabular}
\end{table*}

\section{Conclusions}

This preliminary work motivates further exploration of more sophisticated techniques for pruning Horn clauses in link prediction. Beyond that, we 
can experiment different metrics, such as combinations of the PCA confidence,
which measures the quality of a rule when it is triggered,
with another metric that
can indicate whether the rule is relevant in describing the KG, such as the
support metric.

An interesting direction for future research is also the integration of hybrid techniques, in which heuristics are combined with combinatorial optimisation approaches for pruning and reranking rules. To this end, we plan to investigate the use of MaxSAT~\cite{li2009maxsat} and integer linear programming (ILP) solvers~\cite{gurobi}. 

A similar investigation can be carried out with respect to temporal rule
learners such as TLogic \cite{liu2022tlogic}, which come with a similar 
anytime bottom-up algorithm for learning rules from knowledge graphs.
Beyond temporal rules \emph{forecasting} approaches, where predictions about relations
at timestamp $t$ have access to the KG portion consisting of timestamps $t'<t$ only
\cite{han2021explainable}, some temporal rule approaches have been designed for the 
\emph{interpolation} problem, where rule bodies have access to both past and future information
centered around the timestamp of the query \cite{wang2024mtlearn,solda2026incremental}. This variety of tools may also 
benefit from an adequate post-processing pruning.

\subsubsection*{Acknowledgments}
The authors acknowledge the support by the European Union's Horizon 2020 research and innovation programme
under the Maria Skłodowska-Curie grant agreement
No.~101034440~\marginpar{\worldflag[width=16pt]{EU}}.

\subsubsection*{Use of AI assistants}
We used AI tools to draft and edit text and figures. 
We reviewed all AI-assisted work. Code is checked by tests. 
We take responsibility for the final content of this paper, 
including text, claims and artifacts produced with the aid of generative AI.

\bibliographystyle{plainnat}
\bibliography{references}
\end{document}